\documentclass[prl,twocolumn,floatfix,amsmath,amssymb,superscriptaddress]{revtex4}
\usepackage{graphicx}

\newcommand{\ket}[1]{\ensuremath{\left|{#1}\right\rangle}}

\begin{document}

\title{Efficiency of structured adiabatic quantum computation}

\author{Juan Jos\'e Garc{\'\i}a-Ripoll}
\affiliation{Insituto de F{\'\i}sica Fundamental, CSIC,
  c/Serrano 113b, Madrid 28009, Spain}
\author{Mari Carmen Ba\~nuls}
\affiliation{Max-Planck Institute of Quantum Optics,
  Hans-Kopfermann-Str. 1, Garching, 85748, Germany}

\begin{abstract}
  We show enough evidence that a structured version of Adiabatic
  Quantum Computation (AQC) is efficient for most satisfiability
  problems. More precisely, when the success probability is fixed
  beforehand, the computational resources grow subexponentially in
  the number of qubits. Our study focuses on random satisfiability and
  exact cover problems, developing a multi-step algorithm that solves
  clauses one by one. Relating the computational cost to classical
  properties of the problem, we collect significant statistics with up
  to $N=140$ qubits, around the phase transitions, which is where the
  hardest problems appear.
\end{abstract}

\maketitle

While the general framework of Quantum Computation is to a large
extent justified by the progressive miniaturization of integrated
circuits, there is still a major open problem in the field which is
the finding of new quantum algorithms. This task is rather difficult
due to both the lack of a fully quantum programming model and the
general requirement that any quantum algorithm should outperform its
classical counterparts --a perhaps too ambitious expectation which
imposes both proofs of the classical and quantum complexities.

The Adiabatic Quantum Computation paradigm \cite{farhi01} is a very
appealing framework for the development of new algorithms. While
equivalent to the circuit model~\cite{aharonov04}, this paradigm aims
at adiabatically preparing the ground state of a quantum Hamiltonian,
a state which encodes the solution to a problem we want to solve. The
resource that measures the computational cost in AQC is the time
required to prepare the state and the adiabatic theorem~\cite{messiah}
provides a criterion to bound this time, either numerically or
analytically.  In addition to this, the adiabatic formulation is
particurlarly well suited for constraint solving problems. The
constraints may be encoded as energy penalties in a Hamiltonian, so
that any solution of the whole problem must be a ground state of the
final Hamiltonian.

The traditional interest on AQC is precisely rooted on its relation to
constraint solving and in particular to logical satisfiability (SAT)
problems. From a practical point of view, many real-world problems,
including automated hardward design and verification, can be suitably
represented in SAT form. Furthermore, the subset of 3SAT problems were
also the first problems shown to be
NP-complete~\cite{cook71,levin73}. This implies that if we had an
algorithm to efficiently solve 3SAT, we could also solve all problems
in the much larger NP family~\footnote{These are problems the solution
  of which can be verified in polynomial time}. However, to date, the
extensive literature about classical SAT algorithms suggests that
these problems are exponentially hard to solve~\cite{skjernaa04}, even
in the average and typical cases~\cite{coarfa03}.

Regarding the utility of quantum computation in solving SAT problems,
there is not yet a clear cut answer. Despite the promising results of
the very first AQC simulations~\cite{farhi01}, nowadays AQC is
expected to be expontentially costly for the worst-case instances of
NP-complete SAT
problems~\cite{vandam01,ambainis04,aaronson06,vznindari05}, when no
use of the structure of the problem is made. One may argue that the
worst-case complexity is less relevant than the average-case
cost~\cite{levin84}, and there are evidences for average exponential
speedups in the quantum query problem~\cite{ambainis01}. Many works
have adopted this point of view and studied the running times of
average or typical adiabatic
computations~\cite{farhi01,hogg03,latorre04,banuls06,schutzhold2006,young08}. Here
the results are still mixed, ranging from polynomial~\cite{young08} to
subexponential~\cite{farhi01,banuls06} and to
exponential~\cite{hogg03} growth. These discrepancies can be
attributed to the difficulty of simulating a quantum algorithm, which
limits the analysis to small instances and statistical samples.

In this work we want to rule out not only the worst-case restriction,
but also the use of unstructured algorithms. We will introduce a new
composite algorithm for SAT that alternates diabatic and adiabatic
stages.  We will then relate the running times of the algorithm to
classical properties of the SAT instances. Using this connection, we
exactly characterize the performance of the algorithm for very large
problems, with a much larger sampling space than any previous study
---up to $N=140$ bits with $2\times10^6$ instances---. With this data
we will answer a question of great practical relevance: given a
desired success probability, what is the scaling of the resources
required to solve that fraction of randomly picked, but classically
hard problems. The result will be that the resources grow
subexponentially, even in regimes in which the classical algorithms
would be exponentially costly~\cite{coarfa03}.

\begin{figure}[t]
  \centering
  \includegraphics[width=0.9\linewidth]{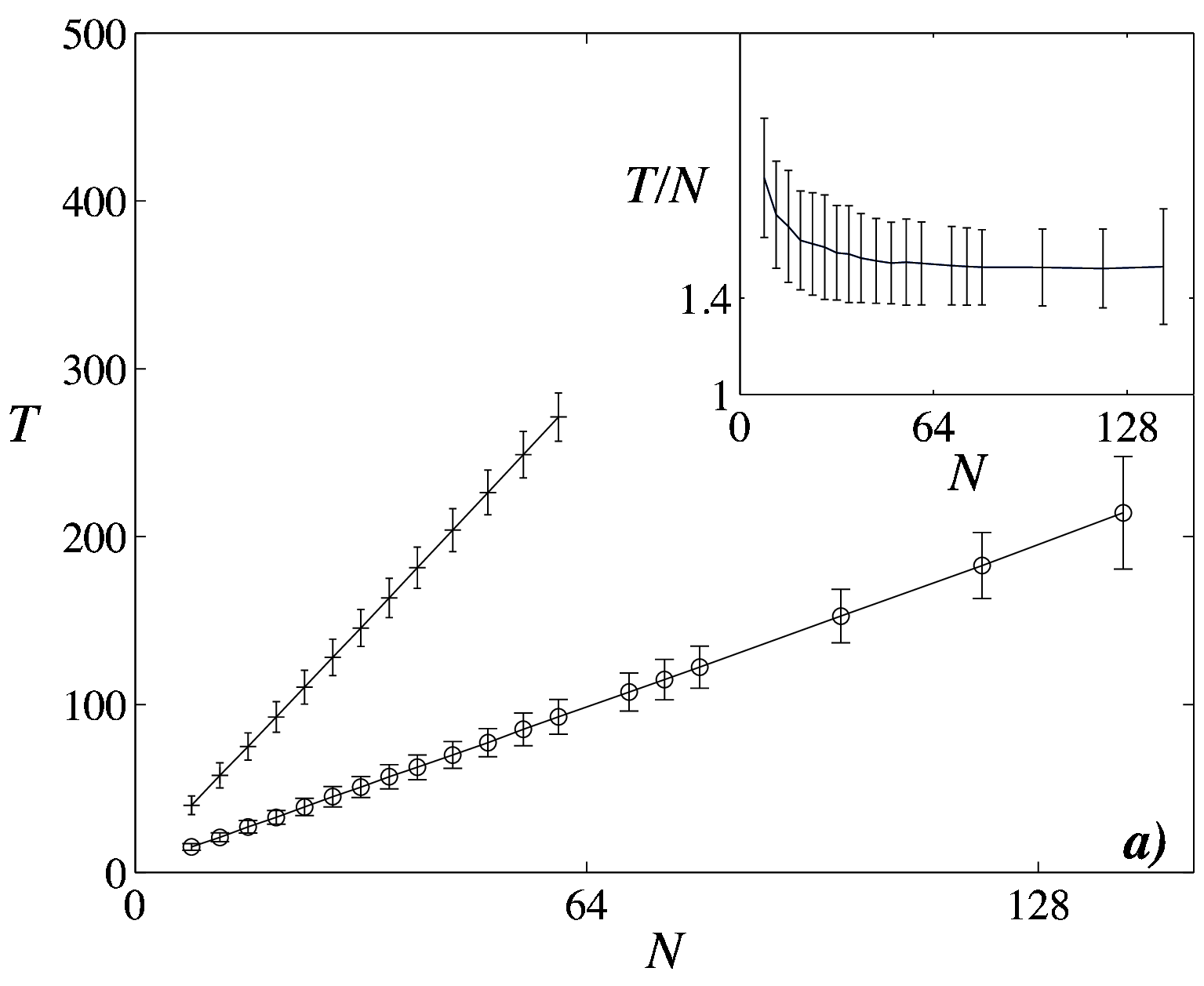}
  \includegraphics[width=0.9\linewidth]{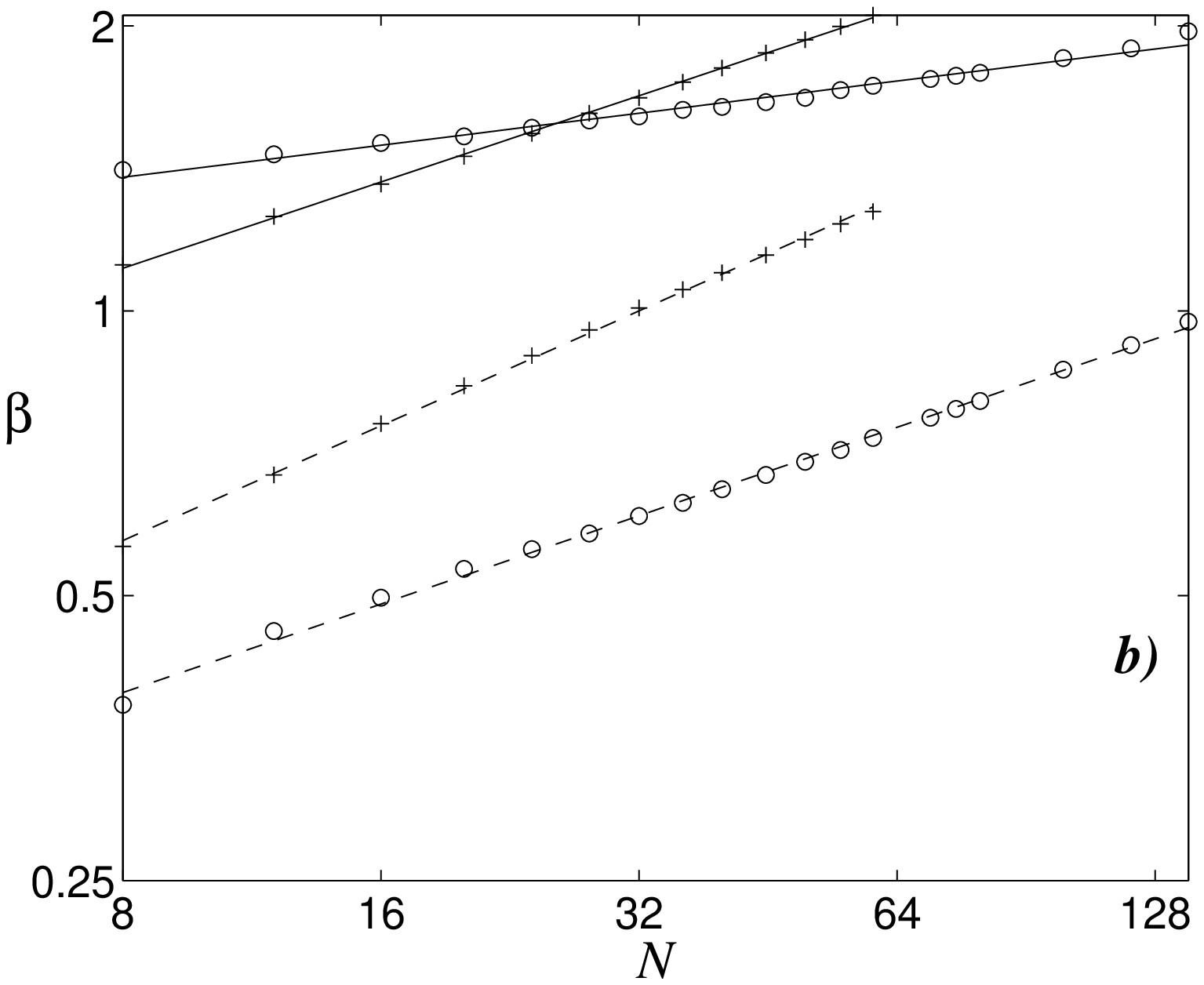}
  \caption{Efficiency of the quantum algorithm. In (a) we plot the
    total running time vs. the number of qubits. We have used
    Eq.~(\ref{time}) with $\Omega=1.$ We plot average results for
    X3SAT (circles) and 3SAT (crosses) together with standard
    deviations. The inset shows the deviation from a linear law for
    X3SAT. Figure (b) is a log-log plot of the average value (solid)
    and variance (dashed) of the value $\beta,$ which is the logarithm
    of the maximum time required by one of the adiabatic
    intervals. The lines are fits to power laws,
    $\beta,\Delta\beta\sim a N^b.$ Each point in the plots involves
    $2\times 10^6$ random instances.}
  \label{fig:time}
\end{figure}

This work focuses on the solution of SAT problems. An instance of this
family is defined by a set of $M$ logical constraints or clauses on a
finite set of $N$ Boolean variables $(s_1,s_2\ldots s_N).$ In the
particular case of $p$-SAT, each clause is a Boolean function
depending only on $p$ variables $ f_k(s_{k_1},s_{k_2},\ldots s_{k_p})
: \{0,1\}^{\otimes p} \to \{0,1\}.$ The problem is to determine
whether there exists an assignment of the $N$ variables satisfying all
$M$ clauses, a task for which there exist both heuristic and
probabilistic classical algorithms.

Our quantum algorithm for SAT is a hybrid that alternates
non-adiabatic evolution with adiabatic steps, in a way that goes
beyond structured AQC~\cite{roland03}. The key ingredients are to
initially sort the set of logical clauses, and to have each adiabatic
step solving a single clause. Let us call $S_m$ the set of assignments
which are compatible with the $m$ first logical clauses. As shown
below, during the $m$-th time interval of adiabatic evolution
$(t_{m-1},t_m),$ the quantum register will evolve from a linear
superposition of all states in $S_{m-1}$ to a linear superposition of
those in $S_m.$ For this step of the computation to succeed we require
a run time
\begin{equation}
  t_m - t_{m-1} = \Omega \times {d_{m-1}}/{d_m},
  \label{time}
\end{equation}
where $d_m=|S_m|$ is the size of the set $S_m$ and $\Omega$ is some
big multiplicative constant specified by the adiabatic
theorem~\cite{messiah}. This relates the resource of the adiabatic
algorithm ---time--- to a classical property ---the structure of
solution spaces---, and it will be the main tool in our analysis of
the efficiency. Note that if we add all clauses in the same adiabatic
interval, the adiabatic time becomes exponentially large
$T_{\mathrm{search}} \sim \Omega d_0/{d_M} \sim 2^{N},$ consistently
with search algorithms for unstructured problems~\cite{roland02}. Let
us now present the algorithm and derive Eq.~(\ref{time}).

\paragraph{The algorithm.}
In AQC, the solution of a computational problem is encoded in a state
$\psi_g$ which is the ground state of a problem Hamiltonian, $H_P$. To
prepare $\psi_g,$ the Hamiltonian of the quantum
register is slowly modulated from an initial operator $H(0)$ with an
easily prepared ground state $\psi(0)$, up to our final operator
$H(T)=H_P.$ The adiabatic theorem states that if the evolution is slow
enough, the quantum register will end up in the desired state,
$\psi(T)=\psi_g.$ The condition ``slow enough'' evolution means that
the changes in the Hamiltonian happen in a long time scale $T\gg
1/(\Delta E)^2,$ larger than the minimal gap $\Delta E$ between the
ground state manifold and the excited states of $H(t)$.

In our algorithm the initial state is a linear combination of all
states in the computational basis $\ket{\psi(0)} \propto
\sum_{\mathbf{s}} \ket{\mathbf{s}} = \left(\ket{0} +
  \ket{1}\right)^{\otimes N}.$ This product state is evolved with a
Hamiltonian that is piecewise continuous. Within the $m$-th interval,
$t \in\left(t_{m-1},t_m\right),$ we write
\begin{equation}
  H(t) = H_C^{(m)} + [1 - \lambda_m(t)] H_B^{(m)} + \lambda_m(t) H_P^{(m)},
  \label{model}
\end{equation}
with adiabatic parameter $\lambda_m(t) = (t - t_{m-1})/(t_m-t_{m-1})$.
The three operators are
\begin{eqnarray}
  H_P^{(m)}\ket{\mathbf{s}} &=& \left[1 - f_m(s_{m_1},s_{m_2},s_{m_3})\right]
  \ket{\mathbf{s}},
  \label{HP}\\
  H_B^{(m)}\ket{\mathbf{s}} &=& -\frac{1}{2^{N/2}}\sum_{\mathbf{s}'} \ket{\mathbf{s}'},\;\mathrm{and}
  \label{HB}\\
  H_C^{(m)} &=& \Gamma \sum_{k=1}^{m-1} H_P^{(k)}.
\end{eqnarray}
Respectively, a term that penalizes invalid configurations for the
$m$-th clause, a component that mixes all possible configurations and
an operator that penalizes any violation of the $m-1$ clauses we have
solved so far. We expect that, if the adiabatic steps succeed and the
non-adiabatic steps do not introduce significant errors, after each
time $t_m$ the register will be in a linear combination of all
assignments which are compatible with the $m$ first logical clauses,
$\psi(t_m)\simeq \ket{\Xi_m} \propto \sum_{s\in S_m}\ket{s}.$

\paragraph{Non-adiabatic errors.}

In between adiabatic intervals the Hamiltonian changes
non-adiabatically from $H(t_m^-)$ to $H(t_m^+).$ This implies
strengthening the last clause by a factor $\Gamma$ and switching on
the term $H_B^{(m)}.$ Without affecting the execution time, this
process introduces a controllable error on the state, which amounts to
the difference between the ground states of $H(t)$, that is
$\psi_g(t),$ immediately before $t_m$ and after it.  Applying the
results in Ref.~\cite{aharonov04} we obtain $\left\langle
  \psi_g(t_m^+) | \psi_g(t^-_m)\right\rangle \leq 1 - \frac{1}{\Gamma
  - 1}.$ Since the difference is ${\cal O}(1/\Gamma),$ the $M$
adiabatic steps will amount to an overall error $M/\Gamma,$ which can
be decreased with resources that are polynomial in $M.$

\paragraph{Cost of the algorithm.}

Given the previous assumption that the constraints term is dominant,
$\Gamma\gg 1,$ we may diagonalize the model Hamiltonian (\ref{model})
approximately, and thus compute the minimum energy gap during each
adiabatic step.  We separate the Hilbert space into two parts, ${\cal
  H}={\cal H}_{m-1} \oplus {\cal H}^\perp_{m-1},$ the space of
solutions which are compatible with the previous $m-1$ clauses, ${\cal
  H}_{m-1}=\mathrm{lin}\{\ket{\bf s},{\bf s} \in S_{m-1}\},$ and its
orthogonal complement. During each interval $(t_{m-1},t_m),$ the low
energy spectrum of $H(t)$ is approximately given~\cite{aharonov04} by
that of its projection onto ${\cal H}_{m-1},$ an operator that we call
$H_{m-1}(t).$ In the two dimensional space spanned by the vector
$\Xi_m$ and its complement $\ket{\Xi^\perp_m} \propto \sum_{\mathbf{s}
  \in {\bar S}_{m}\cap S_{m-1}} \ket{\mathbf{s}},$ we may write
$H_{m-1} = (\lambda_n-\frac{1}{2})\mathbb{I}
+\frac{1}{2}(\lambda-1)\sin(2\alpha_n)\sigma^x
+\frac{1}{2}\left[(\lambda_n-1)\cos(2\alpha_n) -
  \lambda\right]\sigma^z,$ expressed using the Pauli matrices and
$\cos(\alpha_n) = \sqrt{d_n/d_{n-1}} =
\left\langle\Xi_m|\Xi_{m-1}\right\rangle$. This matrix has a minimal
separation or energy gap at $\lambda_n=1/2,$ given by $\Delta E_m =
\cos(\alpha_m) = \sqrt{d_m/d_{m-1}}.$ This provides a first estimate
for the running time of the adiabatic step, $t_m-t_{m-1}\gg 1/(\Delta
E_m)^2$ or Eq.~(\ref{time}). But for this reasoning to be complete, we
must also consider the influence of states with energy $\gg \Gamma.$
Following Ref.~\cite{aharonov04} the high energy states produce
corrections in the energy gap and the ground state wavefunctions which
are small, ${\cal O}(1/\Gamma)$.  Therefore, by imposing a penalty $\Gamma
\simeq \kappa \sqrt{d_{n-1}/d_n}$ at each step, with a constant
$\kappa\gg 1$, or using a global $\Gamma$ larger than all the
individual factors, we can ensure that Eq.~(\ref{time}) still holds.
Finally, since multiplicative factors in the Hamiltonian affect the
total running time, one should actually consider an adimensional
running time~\cite{aharonov04} such as $\tilde T = T \max_s \| H(s)
\|\sim T \times \Gamma,$ but with the previous choices,
both $T$ and $\tilde T$ scale similarly.

\paragraph{Performance.}

For our benchmarks we have chosen two sets of problems: Exact Cover
and 3SAT in conjunctive normal form (CNF). The first family, also
known as X3SAT (or \emph{one-in-three} SAT), has clauses that require
exactly one bit to be set: $f_k(s_{k_1},s_{k_2},s_{k_3})=1$ iff
$s_{k_1}+s_{k_2}+s_{k_3} = 1.$ These problems have been extensively
used to test the efficiency of unstructured adiabatic quantum
computation\cite{farhi01,latorre04,childs05,banuls06}. However, it is
known that even though they form an NP-complete set, Exact Cover
problems tend to be easier than general 3SAT ones. Hence, we have also
studied this more general set of problems, casting them in a
convenient form: $f_k(s_{k_1},s_{k_2},s_{k_3})=0$ iff
$s_{k_1}=a_k,\,s_{k_2}=b_k,\,s_{k_3}=c_k,$ where each function forbids
a precise combination $(a_k,b_k,c_k)$ of the bits involved.

Our statistical study is based on a random sampling of 3SAT and X3SAT
problems, generated with a bias towards hard instances. This is done
by fixing the ratio of clauses to bits, $\alpha=M/N,$ close to the
phase transition from trivially satisfiable problems to problems with
no solution. It is known that the density of hard problems is highest
around these narrow regions
\cite{mitchell92,achlioptas00,achlioptas05,kalapala05,raymond07},
$\alpha_{SAT} \in (4,4.24)$ and $\alpha_{X3SAT}\in(0.54, 0.64),$ while
the outer regions can be easily solved using specialized solvers.  For
each instance we generate $M$ clauses, each one acting on three
different random bits. In the case of 3SAT we also generate the random
bit sequences to be rejected, $(a_k,b_k,c_k)$. Our pseudo-random
number generator is a Mersenne-Twister~\cite{matsumoto98} which has a
period $\sim 2^{19937}$ and good equidistribution properties.

We have applied our algorithm to a sufficiently large selection of
randomly generated 3SAT and Exact Cover instances.
For each random instance of 3SAT or X3SAT the clauses are sorted
according to the indices of the bits that are involved. We loop over
the sorted clauses, iteratively creating the space of valid
assignments, which is decomposed into configurations of the $n_a(m)$
active and $N-n_a(m)$ inactive bits, $S_m=A_m\otimes I_m.$ Since
$d_m=|A_m| 2^{N-n_a(m)}$, only $A_m$ needs to be stored and
computed. The set $A_m$ itself is created from $A_{m-1}$ using a
parallelized algorithm which involves 128 processors and a total of
256Gb of memory for the largest problems. Note that sorting the
clauses is an essential ingredient to make both $n_a(m)$ and $|A_m|$
grow slowly.

The optimal total running time $t_M$ is computed as the sum over all
clauses of the terms in Eq.~(\ref{time}). As shown in
Fig.~\ref{fig:time}a it has an average behavior which is close to
linear in the number of bits. However, preparing an experiment that
runs in this optimal time requires some knowledge about the solution
spaces, as some intervals require more time than others.  A more
realistic goal is choosing a running time that ensures a high
probability of success for arbitrary instances. For this we introduce
a new variable
\begin{equation}
  \beta = \max_m \log_2(d_{m-1}/d_m),\label{beta}
\end{equation}
with which we can uniformly bound the total running time of a given
problem as
\begin{equation}
  t_M \leq \Omega M \times 2^\beta.
\end{equation}
We have computed the value $\beta$ for every instance in the samples. 
By studying the
statistics of this quantity over different problems we infer an
optimal value, $\beta_{opt}(\sigma,N,M),$ such that running the
quantum computer for a time $T_{opt} = \Omega M \exp(\beta_{opt})$
ensures the correct solution of a large fraction $\sigma$ (say
$\sigma=99.99\%$) of all $N$-bit hard instances.  The scaling of
$\beta_{opt}$ with respect to the problem size and success probability
is a more meaningful characterization of the algorithm than the
scaling of the worst case times.

The first results are shown in Fig.~\ref{fig:time}b, where we plot the
average value of $\beta$ and its variance, and fit them to curves of
the form $\beta_{X3SAT} = 1.14 N^{0.162},$ $\Delta\beta_{X3SAT} = 0.1
N^{0.447},$ and $\beta_{3SAT} = 0.468 N^{0.438},$ $\Delta\beta_{3SAT}
= 0.133 N^{0.581}.$ Based on this, the running time without previous
knowledge of the problem could be
\begin{equation}
  T_{opt} \sim \Omega \times N \times 2^{\beta + 6\Delta \beta}
  \sim N 2^{c N^r},
\end{equation}
where exponent $6\Delta\beta$ guarantees a very small probability of
failure. This suggests an overall behavior that follows a
sub-exponential law with $r < 0.5$ for X3SAT and $r<0.6$ for 3SAT.

\begin{figure}
  \centering
  \includegraphics[width=0.9\linewidth]{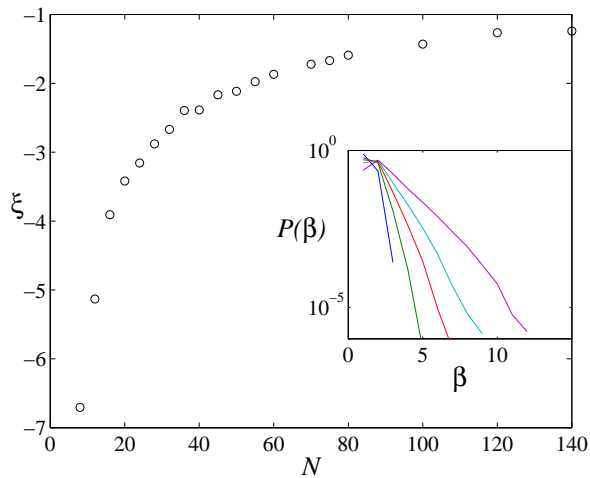}
  \caption{Probability distribution of algorithm running times. The
    inset shows the probability distributions, $P(\beta;N),$ of the
    logarithm of the interval that takes the most time in our
    algorithm (\ref{beta}), for $N=8,16,32,60$ and $140$ bits (left to
    right). These distributions are subject to an exponential fit
    $P(\beta;N) \sim \exp(-\beta/\xi),$ which gives the parameter
    $\xi(N)$ shown in the plot. This value grows subexponentially in
    the number of qubits and is a realistic measure of the algorithm
    efficiency.}
  \label{fig:Pbeta}
\end{figure}

By studying the actual probability distribution of $\beta$ we can
confirm the sub-exponential behavior, obtain a better choice of the
running time and actually estimate the failure probability. Since
$\beta$ cannot be larger than the number of bits, these distributions
have all finite support.  Nevertheless, as shown in
Fig.~\ref{fig:Pbeta}, their tails can be accurately bounded by an
exponential law $\exp(-\beta/\xi).$ According to this law, the failure
probability of a particular setup which uses a running time
$T_{opt}=\Omega M\exp(\beta_{opt})$ is bounded by
\begin{equation}
  P_{\rm fail}=\int_{\beta_{opt}}^{\infty} P(\beta)d\beta \sim
\exp(-\beta_{opt}/\xi).
\end{equation}
Our numerical analysis reveals that $\xi$ grows algebraically with the
number of bits. Therefore, a fixed success probability can be achieved
with subexponential resources.

Summing up, in this work we have presented a hybrid quantum algorithm
which has the advantage that its efficiency can be computed at a
relatively low cost. We have shown that it is possible to establish a
running time such that the algorithm solves a large fraction of
randomly picked but still classically hard problems. This time is
found to scale subexponentially in the size of the problem, which is
to be compared with the average exponential behavior of classical
algorithms~\cite{coarfa03}. While the long range interactions make
this algorithm suboptimal for implementation, our work opens the path
to the development and analysis of other structured algorithms.

JJGR acknowledges support from Spanish projects FIS2006-04885 and
CAM-UCM/910758. The computations for this work were performed at
the Barcelona Supercomputing Center of Spain (BSC-CNS).


\begin{thebibliography}{27}
\expandafter\ifx\csname natexlab\endcsname\relax\def\natexlab#1{#1}\fi
\expandafter\ifx\csname bibnamefont\endcsname\relax
  \def\bibnamefont#1{#1}\fi
\expandafter\ifx\csname bibfnamefont\endcsname\relax
  \def\bibfnamefont#1{#1}\fi
\expandafter\ifx\csname citenamefont\endcsname\relax
  \def\citenamefont#1{#1}\fi
\expandafter\ifx\csname url\endcsname\relax
  \def\url#1{\texttt{#1}}\fi
\expandafter\ifx\csname urlprefix\endcsname\relax\def\urlprefix{URL }\fi
\providecommand{\bibinfo}[2]{#2}
\providecommand{\eprint}[2][]{\url{#2}}

\bibitem[{\citenamefont{Farhi et~al.}(2001)\citenamefont{Farhi, Goldstone,
  Gutmann, Lapan, Lundgren, and Preda}}]{farhi01}
\bibinfo{author}{\bibfnamefont{E.}~\bibnamefont{Farhi}},
  \bibinfo{author}{\bibfnamefont{J.}~\bibnamefont{Goldstone}},
  \bibinfo{author}{\bibfnamefont{S.}~\bibnamefont{Gutmann}},
  \bibinfo{author}{\bibfnamefont{J.}~\bibnamefont{Lapan}},
  \bibinfo{author}{\bibfnamefont{A.}~\bibnamefont{Lundgren}}, \bibnamefont{and}
  \bibinfo{author}{\bibfnamefont{D.}~\bibnamefont{Preda}},
  \bibinfo{journal}{Science} \textbf{\bibinfo{volume}{292}},
  \bibinfo{pages}{472} (\bibinfo{year}{2001}).

\bibitem[{\citenamefont{Aharonov et~al.}(2004)\citenamefont{Aharonov, van Dam,
  Kempe, Landau, Lloyd, and Regev}}]{aharonov04}
\bibinfo{author}{\bibfnamefont{D.}~\bibnamefont{Aharonov}},
  \bibinfo{author}{\bibfnamefont{W.}~\bibnamefont{van Dam}},
  \bibinfo{author}{\bibfnamefont{J.}~\bibnamefont{Kempe}},
  \bibinfo{author}{\bibfnamefont{Z.}~\bibnamefont{Landau}},
  \bibinfo{author}{\bibfnamefont{S.}~\bibnamefont{Lloyd}}, \bibnamefont{and}
  \bibinfo{author}{\bibfnamefont{O.}~\bibnamefont{Regev}},
  \bibinfo{journal}{Foundations of Computer Science, Annual IEEE Symposium on}
  \textbf{\bibinfo{volume}{0}}, \bibinfo{pages}{42} (\bibinfo{year}{2004}).

\bibitem[{\citenamefont{Messiah}(1999)}]{messiah}
\bibinfo{author}{\bibfnamefont{A.}~\bibnamefont{Messiah}},
  \emph{\bibinfo{title}{Quantum Mechanics}} (\bibinfo{publisher}{Dover
  Publications}, \bibinfo{year}{1999}), chap. \bibinfo{chapter}{xvii}.

\bibitem[{\citenamefont{Cook}(1971)}]{cook71}
\bibinfo{author}{\bibfnamefont{S.~A.} \bibnamefont{Cook}}, in
  \emph{\bibinfo{booktitle}{STOC '71: Proceedings of the third annual ACM
  symposium on Theory of computing}} (\bibinfo{publisher}{ACM},
  \bibinfo{address}{New York, NY, USA}, \bibinfo{year}{1971}), pp.
  \bibinfo{pages}{151--158}.

\bibitem[{\citenamefont{Levin}(1973)}]{levin73}
\bibinfo{author}{\bibfnamefont{L.~A.} \bibnamefont{Levin}},
  \bibinfo{journal}{Problems of information transmission}
  \textbf{\bibinfo{volume}{9}}, \bibinfo{pages}{265} (\bibinfo{year}{1973}).

\bibitem[{\citenamefont{Skjernaa}(2004)}]{skjernaa04}
\bibinfo{author}{\bibfnamefont{B.}~\bibnamefont{Skjernaa}}, Ph.D. thesis,
  \bibinfo{school}{University of Aarhus} (\bibinfo{year}{2004}).

\bibitem[{\citenamefont{Coarfa et~al.}(2003)\citenamefont{Coarfa, Demopoulos,
  Aguirre, Subramanian, and Vardi}}]{coarfa03}
\bibinfo{author}{\bibfnamefont{C.}~\bibnamefont{Coarfa}},
  \bibinfo{author}{\bibfnamefont{D.~D.} \bibnamefont{Demopoulos}},
  \bibinfo{author}{\bibfnamefont{A.~S.~M.} \bibnamefont{Aguirre}},
  \bibinfo{author}{\bibfnamefont{D.}~\bibnamefont{Subramanian}},
  \bibnamefont{and} \bibinfo{author}{\bibfnamefont{M.~Y.} \bibnamefont{Vardi}},
  \bibinfo{journal}{Constraints} \textbf{\bibinfo{volume}{8}},
  \bibinfo{pages}{243} (\bibinfo{year}{2003}).

\bibitem[{\citenamefont{van Dam et~al.}(2001)\citenamefont{van Dam, Mosca, and
  Vazirani}}]{vandam01}
\bibinfo{author}{\bibfnamefont{W.}~\bibnamefont{van Dam}},
  \bibinfo{author}{\bibfnamefont{M.}~\bibnamefont{Mosca}}, \bibnamefont{and}
  \bibinfo{author}{\bibfnamefont{U.}~\bibnamefont{Vazirani}},
  \bibinfo{journal}{Foundations of Computer Science, 2001. Proceedings. 42nd
  IEEE Symposium on} pp. \bibinfo{pages}{279--287} (\bibinfo{year}{2001}).

\bibitem[{\citenamefont{Ambainis}(2004)}]{ambainis04}
\bibinfo{author}{\bibfnamefont{A.}~\bibnamefont{Ambainis}},
  \bibinfo{journal}{SIGACT News} \textbf{\bibinfo{volume}{35}},
  \bibinfo{pages}{22} (\bibinfo{year}{2004}).

\bibitem[{\citenamefont{Aaronson}(2006)}]{aaronson06}
\bibinfo{author}{\bibfnamefont{S.}~\bibnamefont{Aaronson}},
  \bibinfo{journal}{SIAM Journal on Computing} \textbf{\bibinfo{volume}{35}},
  \bibinfo{pages}{804} (\bibinfo{year}{2006}).

\bibitem[{\citenamefont{\v{Z}nidari\v{c}}(2005)}]{vznindari05}
\bibinfo{author}{\bibfnamefont{M.}~\bibnamefont{\v{Z}nidari\v{c}}},
  \bibinfo{journal}{Phys. Rev. A} \textbf{\bibinfo{volume}{71}},
  \bibinfo{eid}{062305} (\bibinfo{year}{2005}).

\bibitem[{\citenamefont{Levin}(1986)}]{levin84}
\bibinfo{author}{\bibfnamefont{L.~A.} \bibnamefont{Levin}},
  \bibinfo{journal}{SIAM Journ. Comp.} \textbf{\bibinfo{volume}{15}},
  \bibinfo{pages}{285} (\bibinfo{year}{1986}).

\bibitem[{\citenamefont{Ambainis and de~Wolf}(2001)}]{ambainis01}
\bibinfo{author}{\bibfnamefont{A.}~\bibnamefont{Ambainis}} \bibnamefont{and}
  \bibinfo{author}{\bibfnamefont{R.}~\bibnamefont{de~Wolf}},
  \bibinfo{journal}{J. Phys. A: Math. Gen.} \textbf{\bibinfo{volume}{34}},
  \bibinfo{pages}{6741} (\bibinfo{year}{2001}).

\bibitem[{\citenamefont{Hogg}(2003)}]{hogg03}
\bibinfo{author}{\bibfnamefont{T.}~\bibnamefont{Hogg}}, \bibinfo{journal}{Phys.
  Rev. A} \textbf{\bibinfo{volume}{67}}, \bibinfo{pages}{022314}
  (\bibinfo{year}{2003}).

\bibitem[{\citenamefont{Latorre and Or\'us}(2004)}]{latorre04}
\bibinfo{author}{\bibfnamefont{J.~I.} \bibnamefont{Latorre}} \bibnamefont{and}
  \bibinfo{author}{\bibfnamefont{R.}~\bibnamefont{Or\'us}},
  \bibinfo{journal}{Phys. Rev. A} \textbf{\bibinfo{volume}{69}},
  \bibinfo{pages}{062302} (\bibinfo{year}{2004}).

\bibitem[{\citenamefont{Ba{\~n}uls et~al.}(2006)\citenamefont{Ba{\~n}uls,
  Or\'{u}s, Latorre, P\'{e}rez, and Ruiz-Femen\'{\i}a}}]{banuls06}
\bibinfo{author}{\bibfnamefont{M.~C.} \bibnamefont{Ba{\~n}uls}},
  \bibinfo{author}{\bibfnamefont{R.}~\bibnamefont{Or\'{u}s}},
  \bibinfo{author}{\bibfnamefont{J.~I.} \bibnamefont{Latorre}},
  \bibinfo{author}{\bibfnamefont{A.}~\bibnamefont{P\'{e}rez}},
  \bibnamefont{and}
  \bibinfo{author}{\bibfnamefont{P.}~\bibnamefont{Ruiz-Femen\'{\i}a}},
  \bibinfo{journal}{Phys. Rev. A} \textbf{\bibinfo{volume}{73}},
  \bibinfo{eid}{022344} (\bibinfo{year}{2006}).

\bibitem[{\citenamefont{Schutzhold and Schaller}(2006)}]{schutzhold2006}
\bibinfo{author}{\bibfnamefont{R.}~\bibnamefont{Schutzhold}} \bibnamefont{and}
  \bibinfo{author}{\bibfnamefont{G.}~\bibnamefont{Schaller}},
  \bibinfo{journal}{Phys. Rev. A} \textbf{\bibinfo{volume}{74}}
  (\bibinfo{year}{2006}).

\bibitem[{\citenamefont{Young et~al.}(2008)\citenamefont{Young, Knysh, and
  Smelyanskiy}}]{young08}
\bibinfo{author}{\bibfnamefont{A.~P.} \bibnamefont{Young}},
  \bibinfo{author}{\bibfnamefont{S.}~\bibnamefont{Knysh}}, \bibnamefont{and}
  \bibinfo{author}{\bibfnamefont{V.~N.} \bibnamefont{Smelyanskiy}},
  \bibinfo{journal}{Phys. Rev. Lett.} \textbf{\bibinfo{volume}{101}},
  \bibinfo{pages}{170503} (\bibinfo{year}{2008}),
  \bibinfo{note}{arXiv:0803.3971}.

\bibitem[{\citenamefont{Roland and Cerf}(2003)}]{roland03}
\bibinfo{author}{\bibfnamefont{J.}~\bibnamefont{Roland}} \bibnamefont{and}
  \bibinfo{author}{\bibfnamefont{N.~J.} \bibnamefont{Cerf}},
  \bibinfo{journal}{Phys. Rev. A} \textbf{\bibinfo{volume}{68}},
  \bibinfo{pages}{062312} (\bibinfo{year}{2003}).

\bibitem[{\citenamefont{Roland and Cerf}(2002)}]{roland02}
\bibinfo{author}{\bibfnamefont{J.}~\bibnamefont{Roland}} \bibnamefont{and}
  \bibinfo{author}{\bibfnamefont{N.~J.} \bibnamefont{Cerf}},
  \bibinfo{journal}{Phys. Rev. A} \textbf{\bibinfo{volume}{65}},
  \bibinfo{pages}{042308} (\bibinfo{year}{2002}).

\bibitem[{\citenamefont{Childs et~al.}(2001)\citenamefont{Childs, Farhi, and
  Preskill}}]{childs05}
\bibinfo{author}{\bibfnamefont{A.~M.} \bibnamefont{Childs}},
  \bibinfo{author}{\bibfnamefont{E.}~\bibnamefont{Farhi}}, \bibnamefont{and}
  \bibinfo{author}{\bibfnamefont{J.}~\bibnamefont{Preskill}},
  \bibinfo{journal}{Phys. Rev. A} \textbf{\bibinfo{volume}{65}},
  \bibinfo{pages}{012322} (\bibinfo{year}{2001}).

\bibitem[{\citenamefont{Mitchell et~al.}(1992)\citenamefont{Mitchell, Selman,
  and Levesque}}]{mitchell92}
\bibinfo{author}{\bibfnamefont{D.}~\bibnamefont{Mitchell}},
  \bibinfo{author}{\bibfnamefont{B.}~\bibnamefont{Selman}}, \bibnamefont{and}
  \bibinfo{author}{\bibfnamefont{H.}~\bibnamefont{Levesque}}, in
  \emph{\bibinfo{booktitle}{Proceedings of the Tenth National Conference on
  Artificial Intelligence (AAAI-92), San Jose, CA}} (\bibinfo{year}{1992}), pp.
  \bibinfo{pages}{459--465}.

\bibitem[{\citenamefont{Achlioptas et~al.}(2000)\citenamefont{Achlioptas,
  Gomes, Kautz, and Selman}}]{achlioptas00}
\bibinfo{author}{\bibfnamefont{D.}~\bibnamefont{Achlioptas}},
  \bibinfo{author}{\bibfnamefont{C.}~\bibnamefont{Gomes}},
  \bibinfo{author}{\bibfnamefont{H.}~\bibnamefont{Kautz}}, \bibnamefont{and}
  \bibinfo{author}{\bibfnamefont{B.}~\bibnamefont{Selman}}, in
  \emph{\bibinfo{booktitle}{In AAAI/IAAI}} (\bibinfo{publisher}{AAAI Press},
  \bibinfo{year}{2000}), pp. \bibinfo{pages}{256--261}.

\bibitem[{\citenamefont{Achlioptas et~al.}(2005)\citenamefont{Achlioptas, Naor,
  and Peres}}]{achlioptas05}
\bibinfo{author}{\bibfnamefont{D.}~\bibnamefont{Achlioptas}},
  \bibinfo{author}{\bibfnamefont{A.}~\bibnamefont{Naor}}, \bibnamefont{and}
  \bibinfo{author}{\bibfnamefont{Y.}~\bibnamefont{Peres}},
  \bibinfo{journal}{Nature} \textbf{\bibinfo{volume}{435}},
  \bibinfo{pages}{759} (\bibinfo{year}{2005}).

\bibitem[{\citenamefont{{Kalapala} and {Moore}}()}]{kalapala05}
\bibinfo{author}{\bibfnamefont{V.}~\bibnamefont{{Kalapala}}} \bibnamefont{and}
  \bibinfo{author}{\bibfnamefont{C.}~\bibnamefont{{Moore}}},
  \bibinfo{note}{arXiv:cs/0508037}.

\bibitem[{\citenamefont{Raymond et~al.}(2007)\citenamefont{Raymond, Sportiello,
  and Zdeborova}}]{raymond07}
\bibinfo{author}{\bibfnamefont{J.}~\bibnamefont{Raymond}},
  \bibinfo{author}{\bibfnamefont{A.}~\bibnamefont{Sportiello}},
  \bibnamefont{and}
  \bibinfo{author}{\bibfnamefont{L.}~\bibnamefont{Zdeborova}},
  \bibinfo{journal}{Phys. Rev. E} \textbf{\bibinfo{volume}{76}},
  \bibinfo{pages}{011101} (\bibinfo{year}{2007}).

\bibitem[{\citenamefont{Matsumoto and Nishimura}(1998)}]{matsumoto98}
\bibinfo{author}{\bibfnamefont{M.}~\bibnamefont{Matsumoto}} \bibnamefont{and}
  \bibinfo{author}{\bibfnamefont{T.}~\bibnamefont{Nishimura}},
  \bibinfo{journal}{ACM Trans. Model. Comput. Simul.}
  \textbf{\bibinfo{volume}{8}}, \bibinfo{pages}{3} (\bibinfo{year}{1998}).

\end{thebibliography}
\end{document}